\documentclass[prb,twocolumn,showpacs,floatfix]{revtex4}

\usepackage{graphicx}
\usepackage{array}
\usepackage{amsmath,amsfonts,amssymb}
\usepackage[ansinew]{inputenc}

\newcommand{\buck}{C$_{60}$}
\newcommand{\Figref}[1]{Fig.~\ref{#1}}

\begin{document}

\title{Passing current through touching molecules}

\author
{Guillaume Schull,$^1$ Thomas Frederiksen,$^2$ Mads Brandbyge,$^3$ Richard
Berndt$^1$ }

\affiliation{$^1$Institut f\"{u}r Experimentelle und Angewandte
Physik,Christian-Albrechts-Universit\"{a}t zu Kiel, D-24098 Kiel, Germany}
\affiliation{$^2$Donostia International Physics Center (DIPC), E-20018
Donostia-San Sebasti\'an, Spain}
\affiliation{$^3$DTU Nanotech, Technical University of Denmark, DK-2800
Kgs.~Lyngby, Denmark}

\begin{abstract}
The charge flow from a single C$_{60}$\ molecule to another one
has been probed. The conformation and electronic
states of both molecules on the contacting electrodes have been
characterized using a cryogenic scanning tunneling microscope.  While
the contact conductance of a single molecule between two Cu electrodes can vary
up to a factor of three depending on electrode geometry, the conductance
of the C$_{60}$--C$_{60}$\ contact is consistently lower by two orders of
magnitude.
First-principles transport calculations reproduce the experimental results,
allow a determination of the actual C$_{60}$--C$_{60}$\ distances, and identify
the essential
role of the intermolecular link in bi- and trimolecular chains.  
\end{abstract}


\maketitle

Intermolecular charge transport is central to numerous research fields. In
biology electron hopping and tunneling processes between molecules play a vital
role.\cite{Boon2003,Miyashita2005}  Moreover, 
tunneling processes between molecular materials have opened new perspectives
towards
the realization of efficient molecular sensors and solar 
cells.\cite{El-Khouly2004,Goldsmith2005}
In a parallel direction the conductance 
properties of point contacts, \cite{Jansen1980} single atoms \cite{Muller1992}
or single molecules \cite{Joachim1995} are intensely being investigated, and
give a detailed view of charge transport through individual nanoscopic objects.
Recently, experiments realized on 1D extended molecules, \cite{HoChoi2008}
single conjugated polymers \cite{Lafferentz2009} and DNA wires \cite{Cohen2005}
have been reported. A critical issue is now to understand and control the charge
transfer from a single molecule to another one.

Here we probe the electron current passing a chain of two \buck\ molecules
suspended in a STM junction, where the orientation and electronic states of
both molecules have been characterized before connecting them with atomic-scale
precision.  The experimental results are complemented by first-principles
transport calculations which give access to the distance-dependent nature of the
intermolecular electron transport and predict the evolution of the transport
properties with molecular chain length.

The experiments were performed with a low-temperature STM operated at 5.2 K in
ultrahigh vacuum (below 10$^{-8}$ Pa). Au(111) and Cu(111) samples and
etched W tips were prepared by Ar$^+$ bombardment and annealing. As a final
preparation, W tips were indented into the sample surface to coat them
with surface material. \buck\ molecules were
deposited from a Ta crucible at a rate of $\approx 1 $ ML/min as monitored by a
quartz microbalance. During deposition a residual gas pressure in the 10$^{-6}$
Pa range was maintained and the sample was kept at room temperature. The data
shown correspond to a coverage of approximately 0.2 \buck\ monolayers. All
images were recorded in a constant-current mode.

\begin{figure}
  \includegraphics[width=0.95\linewidth]{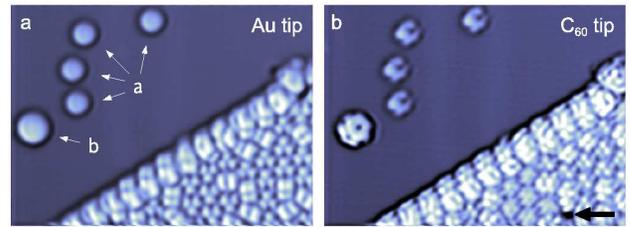}
  \caption{STM images ($I = 10$ nA; $V = 2.5$ V; $14.3\times 10.5$ nm$^2$) of a
Au(111) surface
partially 
covered with \buck\ molecules (lower right) obtained with a (a) metal and (b)
\buck\ 
tip over the same area. Gold adatoms ($\alpha$) and a small gold cluster
($\beta$) of two or 
three adatoms are discernible.}
  \label{fig1}
\end{figure}

Increased image resolution with molecule-covered STM tips has repeatedly been
reported. \cite{Xu2001,Repp2005,Nishino2005}
However, no detailed information about the molecular orientation or their
electronic properties was available.  To realize controlled molecular contacts,
these details are decisive. We therefore used \buck\ molecules as their
orientation can be determined from sub-molecularly
resolved STM images. \cite{Lu2003,Schull2007}  Figure \ref{fig1}a shows a STM
image, recorded with a metallic tip, of an array of \buck\ on a Au(111) surface.
Two \buck\ orientations
are observed, which are typical of a $(2\sqrt{3}\times 2\sqrt{3})$R$30^\circ$
\buck\ superlattice.\cite{Schull2008}

 To attach a \buck\ molecule
to the tip, the metallic tip was placed over a target molecule and the sample
voltage was varied from 2 V to 0.01 V and back at a constant current $I = 100$
nA. The success of this procedure can be verified from the removal of the
molecule
from the substrate (e.\,g., missing \buck\ in \Figref{fig1}b indicated by a black arrow). 
To further characterize \buck\ tips, structures composed of one ($\alpha$) and
two or three ($\beta$) Au adatoms had been deposited by slight contacts of the
metallic STM tip with a clean surface area (\Figref{fig1}a, upper left
corner).\cite{Limot2005}  The image of \Figref{fig1}b was obtained with a
\buck\ functionalized tip over the same area. The Au clusters, which appear
round and featureless with a metallic tip, exhibit a complex pattern which
matches the highest
occupied molecular orbital (HOMO) of \buck.\cite{Lu2003}  Obviously the Au
adatoms work as tips for ''reverse'' imaging of  \buck\ at the tip and provide
direct
access to the orientation of the molecule, e.\,g., in \Figref{fig1}b, a 6:6
bond of the \buck\ tip is facing the surface.  While this
technique has previously been used to determine the number of molecules
adsorbed on a STM tip,\cite{Kelly1996a} the characterization of submolecular 
structures was not reported. 

\begin{figure}
  \includegraphics[width=0.95\linewidth]{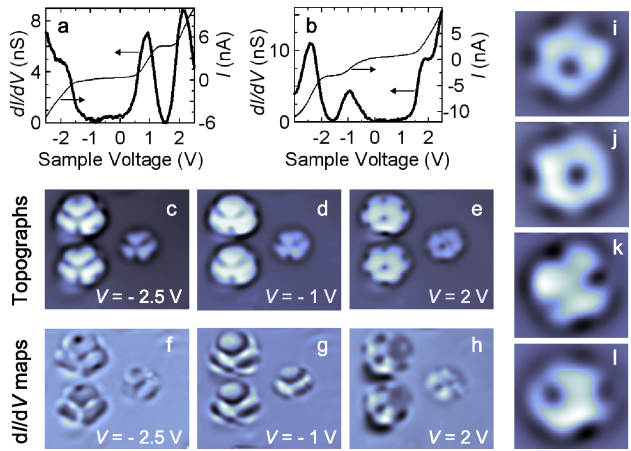}
  \caption{Differential conductance ($dI/dV$) spectra
acquired over (a) a  
\buck\ molecule with a metallic tip and (b) the bare metal with a \buck\ tip
(Set point $I = 10$ nA; $V = 2.5$ V; rms modulation 10~mV).
(c), (d) and (e) are reverse STM images ($I = 10$ nA; $5\times 4.2$ nm$^2$)
acquired over $\alpha$ and $\beta$ type 
atomic clusters with a \buck\ tip at the voltages corresponding to peaks in the
spectra of (b).
(f), (g), and (h) are conductance maps acquired simultaneously
with each picture
(rms modulation 50~mV).  Reverse STM images (i) to (l) of a Au
adatoms with a \buck\ tip ($I = 10$ nA; $V = 2.5$ V; $1.5\times 1.4$ nm$^2$). Between
image acquisitions, the was approached to the surface so as to 
reach $I \approx 1 \mu$A, where reorientation of the molecule at
the tip occurred.}
  \label{fig2}
\end{figure}

\begin{figure}
  \includegraphics[width=0.95\linewidth]{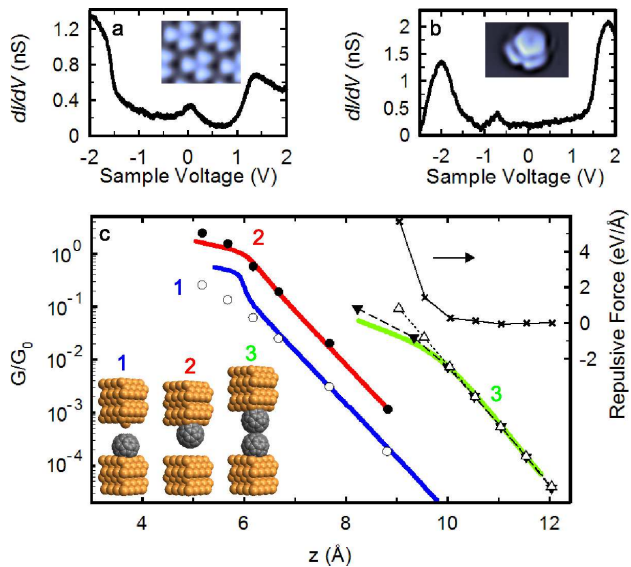}
  \caption{(a) $dI/dV$ spectra
acquired over (a) a  
\buck\ molecule on Cu(111) with a metallic tip and (b) the bare metal with a
\buck\ tip
(Set point $I = 10$ nA; $V = 2$ V; rms modulation 10~mV.
The insets show (a) a typical STM image ($I = 10$ nA; $V = 2$ V; $2\times
1.6$ nm$^2$) of a 
\buck\ array on Cu(111), where all molecules expose hexagons to vacuum, and (b)
a reverse STM image ($I = 10$ nA; $V = -2$ V; $2.7\times 1.8$ nm$^2$) of the
\buck\ tip used for the contact experiment (a 5:6 bond is exposed to the
surface). The inset to (c) displays sketches of the contact experiments
performed by approaching (1) a sharp metallic tip to a \buck\ adsorbed on a
hexagon on Cu(111), a 5:6 oriented \buck\ tip (2) to the bare Cu(111) surface and
(3) to a \buck\ adsorbed on a hexagon. (c)
Experimental (solid lines) and calculated (symbols) conductances of the junctions
versus the tip--molecule distance (1) (Set point $I = 0.1$ nA; $V = 100$ mV) and
(2) (Set point $I = 10$ nA; $V = - 100$ mV) and the
\buck\ to \buck\ center distance (3) (Set point $I = 0.1$ nA; $V = 200$ mV). The
calculated repulsive force (crosses) between two \buck\ molecules
suggests an elastic deformation of the junction at small separations
that maps real molecule--molecule distances
(open triangles) with apparent distances (filled triangles), see text.}
  \label{fig3}
\end{figure}

To monitor the density of states of \buck\ tips, \Figref{fig2} displays
conductance spectra obtained with (a) a metallic tip on \buck\ and (b)
a ''reverse'' spectrum recorded with a \buck\ tip on bare Au.
The spectral peaks are characteristics of the molecular orbitals of \buck\ on Au(111).
\cite{Schull2008} The spectra are almost perfect mirror images of each other
reflecting that the electronic state of \buck\ at the tip is closely related to
those of \buck\ on the surface. ''Reverse'' images and conductance maps of
atomic sized clusters exhibit submolecular patterns
(\Figref{fig2}c to h) which are typical of the lowest unoccupied molecular orbitals
(LUMO+1, $V \approx -2.5$ V and LUMO, $V \approx -1$ V) and HOMO ($V \approx 2$ V).\cite{Lu2003,Schull2008} 
Once a molecule is attached
to the tip it is possible to change its orientation by passing current of up to
$\approx 1 \mu$A as demonstrated in
Figs.\ \ref{fig2}i to l. The molecular patterns obtained correspond to different
\buck\ orientations at the tip. While this sequence
demonstrates control over the orientation of the tip molecule it also
highlights an instability of these tips at high currents, which, therefore, were
not suitable for the intended contact experiments. We repeated the previous
experiments on \buck\ deposited on a Cu(111) substrate (\Figref{fig3}a) where
the binding of \buck\ is stronger. \cite{Schull2008} Transfer of \buck\ from the
Cu(111) to the Cu-covered tip remains feasible, although the procedure is less
reproducible than for Au(111) and $\mu$A-currents are required. As in the Au
case, the structural and electronic properties of the \buck-tips have been
characterized (\Figref{fig3}b).
 
Following their characterization, metal and \buck\ tips 
were approached to \buck\ molecules and pristine Cu(111) areas
and conductance--distance [$G(z)$] data were recorded (Fig.\ \ref{fig3}c).
No reorientation of the molecules occurred. 
Curve 1 was obtained with a sharp metallic tip approaching a \buck\
molecule. The right part of the trace corresponds
to the tunneling range.  Contact is indicated by an inflection of
the trace, which defines a contact conductance of $0.3$ G$_0$
(G$_0 = 2 e^2 / h$ is the conductance quantum) in agreement with previous
measurements on similar systems. \cite{Joachim1995,Neel2007} Curve 2 represents
a measurement with a \buck\ tip approaching pristine Cu.
Surprisingly, the
contact conductance of $1.0\:$G$_0$ is substantially higher
than with \buck\ on the surface. A conductance of $\approx\ $1.5 G$_0$ at the
contact constituted an upper limit which was not exceeded for different \buck\
tip orientations.

To understand the measured conductance traces first-principles transport
simulations were 
carried out for representative junctions containing \buck\ molecules between
Cu(111) electrodes. We modelled the fullerene junctions by supercells with one
or more \buck\ molecules bridging a 4 $\times$ 4 representation of a slab
containing 13 Cu(111) layers. The electronic structure was determined with the
SIESTA pseudopotential density functional theory (DFT) code\cite{Soler.02} to calculate the
transport properties for the TranSIESTA 
setup.\cite{Brandbyge.02} (For details cf.\ Ref.\ \onlinecite{epaps}). Case 1
was modeled
with a \buck\ adsorbed on a hexagon on the substrate side centered 
underneath a Cu adatom on the tip side (cf.\ inset to \Figref{fig3}c), and case 2
with a \buck\
adsorbed on a 5:6 bond on the tip side facing a clean Cu(111) surface.
For a transparent interpretation of the experiments we have considered
the tip--molecule (sample--molecule) separation as the only variable in case 1 (case 2), and full geometry relaxations
were not performed. Except for the structural rearrangements expected with a
sharp metallic tip (case 1),\cite{Neel2007} this 
approach reproduces and explains the observed traces.
The calculated zero-bias conductances (\Figref{fig3}c) enable a calibration
of the absolute distances $z$ (outermost Cu atom to \buck-center along the surface
normal) between tip (sample) and molecule in case 1(2) 
by aligning the tunneling part of the traces. 
Comparison of cases 1 and 2 shows that for a given distance $z$,
depending on the geometry of the molecule-electrode interface, the conductance
of a single \buck\ junction can vary by a factor of 3 (10)
under contact (tunneling) conditions. 
The conductance of the \buck/Cu(111) junctions is dominated by the molecular
LUMO resonances that lie closest to the Fermi energy $E_F$. The theoretical maximum is
therefore 3 G$_0$ corresponding to three fully
open conductance eigenchannels.\cite{Brandbyge.02}
Indeed, a decomposition $T=\sum_iT_i$ 
of the total transmission $T=T(E_F)$ into eigenchannel contributions $\{T_i\}$
confirms that the three most
transmitting channels carry about an order of magnitude more current than the
fourth. For the sharp-tip contact (case 1 in Fig.\ 3) the transmissions
in contact are of the order $\{T_i\}\approx \{0.12, 0.08, 0.04,0.004\}$, hence
the majority of an incoming electron wave is being reflected in this type of
junction. Contrary, for the \buck-tip contact (case 2 in Fig.\ 3), three
channels are much more open, theoretically in one case as much as
$\{T_i\}\approx \{0.97, 0.87, 0.57, 0.02\}$.

\begin{figure}
  \includegraphics[width=0.95\linewidth]{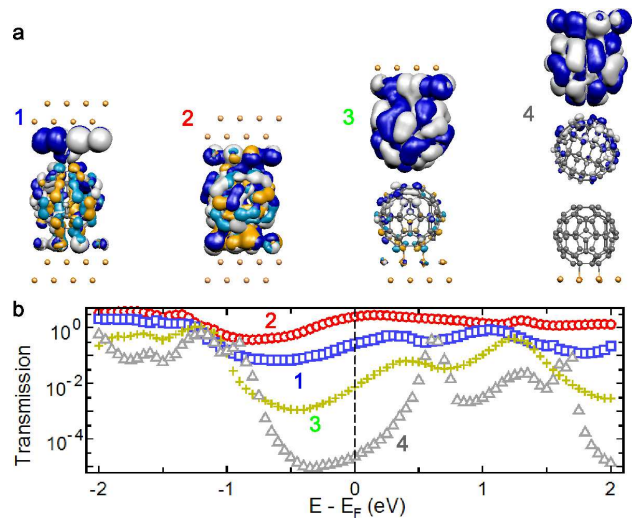}
  \caption{(a) Visualizations of the most transmitting eigenchannel wave function
(incoming from above) around the Fermi energy $E_F$.
The isosurfaces of the real (imaginary) part of the wave 
function (with sign) are colored in white/dark blue (orange/light blue).  
(b) Calculated transmission functions for suspended molecular
chains made of one, two, or three \buck\ molecules. The molecular orientations 
correspond to (1) an adatom vs.~a hexagon, (2) a 5:6 bond vs.~a flat surface,
(3) a hexagon vs.~a 5:6 bond,
and (4) a hexagon vs.~a 5:6 bond vs.~a hexagon.}
  \label{fig4}
\end{figure}

To find out \emph{where} the electrons are being scattered,
\Figref{fig4}a visualizes the most transmitting eigenchannel wave function for
the different contacts.\cite{Paulsson.07}  The isosurfaces
are colored according to the phase as explained
in the figure caption. Since the absolute square of the wave function corresponds to the 
density of the traversing electrons, the magnitude of the lobes
gives an idea where the electron wave travels. In case 1 (sharp tip) the current is
scattered at the single-atom contact to the molecule as indicated
by the standing wave pattern at the tip side. 
In case 2 (\buck-tip) the channel is
almost perfectly open and the wave is propagating
with essentially equal amplitude on either side of the
molecule. To disentangle the effects of different molecular
orientation as well as of different atomic contacts on the conductance, we have carried 
out separate calculations with the hexagon
orientation contacted with a flat tip. Specifically, between
flat Cu(111) electrodes the 5:6 orientation was found to conduct slightly less
than the hexagon orientation, e.g., at $z=6.68$ {\AA} the conductance is
19 {\%} lower for the 5:6 configuration. Moreover, experimental data from various \buck\ orientations \cite{neelnano} show that 0.3 G$_0$ (case 1) is already an upper limit of
the conductance of \buck\ on a Cu surface contacted with a sharp STM tip. We therefore 
conclude that the higher conductance of 1.0 G$_0$ (case 2) is due to the multiple-atomic 
contact which ensures a better connection 
between the molecule and the electrode. This characterization 
of the metal--molecule contact could be valuable for 
fullerene-based anchoring strategies for molecular electronics.\cite{Martin2008}

Finally, the \buck\ tip of Curve 2 in \Figref{fig3}c was also
approached to a \buck\ molecule on the substrate. Yazdani \textit{et al.} used a
similar method to measure the conductance of diatomic Xenon chain.\cite{AliYazdani1996}
Employing the molecular
orientations determined in the experiment, this case was modeled with a \buck\
adsorbed on a hexagon on 
the substrate side under a \buck\ adsorbed on a 5:6 bond on the tip side. Note that the displacement axis now shows the \buck\ to
\buck\ center distance. The observed conductance trace (Curve 3 in
\Figref{fig3}c) varies smoothly from tunneling to contact at 
a molecular separation of $\approx 1$ nm. The contact 
conductance of $\approx 0.01$ G$_0$ is an order of magnitude smaller than
expected for a \buck\ dimer.\cite{Ono2007}
In contrast to the experimental observation of a plateau, our model predicts an
exponential dependence of the conductance on the molecule--molecule 
separation (open triangles in \Figref{fig3}c) and no significant influence of
the
molecule--surface distance (see Ref.\ \onlinecite{epaps}). 
This difference is due to an
intermolecular repulsion at small distances that deforms the
contact. Therefore, beyond the point of contact, the experimental 
data reflect \emph{apparent} molecular separations, the actual distances being
somewhat larger. To take into account the elastic deformation of the junction,
repulsive forces were estimated from our DFT
calculations (crosses in \Figref{fig3}c). By renormalizing the theoretical
z-coordinates according to the compliance of two soft molecule-surface segments
(effective elastic constant $k_\mathrm{eff} = 7$ eV/{\AA}),\cite{epaps} we
obtain a good agreement
with the experimental trace (filled triangles in \Figref{fig3}c). While it
is possible to further improve this agreement by considering the elasticity
of the other parts of the system,\cite{epaps} the
essential
feature of the experiment is already captured by inclusion of just the
softest segment.
Interestingly, the onset of elastic deformation coincides with the
intermolecular distance of 
$1.004$ nm in \buck\ crystals, which is controlled by 
van der Waals bonding and electrostatic repulsion.\cite{David1992}

The picture emerging for chains of two \buck\ molecules is 
that the transport processes are mainly sensitive
to the molecule--molecule interface. It is further supported by
\Figref{fig4}a3 
which shows isosurfaces of the dominant eigenchannel with little weight on the lower
molecule. This is due to a reduced wave function amplitude beyond
the \buck--\buck\ interface, which thus acts as conductance bottleneck. Within the
chain the intermolecular distance is limited by electrostatic repulsion.
In this way the experiment is probing how current passes through two
\emph{touching} molecules,
the properties and the nature of both being controlled and tunable.

Using the characteristic \buck--\buck\ contact distance determined above, the
transport through a three-\buck\ chain was calculated. In this case the
dominant eigenchannel \Figref{fig4}a4 is strongly attenuated along the chain as
revealed by the absence of lobes on the lower molecule.  The
transmission functions of the molecular chains \Figref{fig4}b reveal
the opening of a $\sim1.5$ eV gap around $E_F$, and hence predict a rapid evolution
towards the insulating character of an infinite \buck\ chain. 

In summary, the contact conductance for single \buck\ junctions can vary up to a
factor of three depending on the molecule--metal interfaces, thus corroborating
the notion of good and bad contacts. The current passing from one molecule to
another one, however, is determined by the molecule--molecule interface. Our
experimental approach can be extended to a range of molecules to address the
influence of the molecule--molecule interactions on intermolecular charge
transport. Moreover, through detection of photons emitted in a STM
junction,\cite{Schull2008a} a suitable fluorescent molecule attached to the STM
tip might prove useful as optically active probe.\cite{Michaelis2000}

Financial support by the Deutsche Forschungsgemeinschaft (SFB 677) and the Danish FNU
(272-07-0114) is gratefully acknowledged.

\newpage
\renewcommand{\thefigure}{S\arabic{figure}}
\setcounter{figure}{0}

\section{Supplementary material}

\textbf{Theoretical details.}
The electronic structure for the supercells shown in the inset of Fig.\ 3c was calculated
with the SIESTA\cite{Soler.02.S} pseudopotential density functional theory (DFT) method
and the generalized gradient approximation \cite{Perdew.96.S} for
exchange-correlation (XC). A standard single-zeta plus polarization basis was
employed for Cu (0.15 eV energy shift, $r_c$ $\leq$ 6.9 a.u.) and a long-ranged,
double-zeta plus polarization basis for C (0.02 eV energy shift, $r_c$ $\leq$
5.5 a.u.). Real-space grid integrations were carried out using a 200 Ry energy
cutoff. The 3D Brillouin zone was sampled with a $2\times2\times1$
Monkhorst-Pack $\mathbf k$-mesh. The lattice constant for the Cu crystal was set
to 3.70 {\AA}. An initial structure consisting of a \buck\ molecule adsorbed
with a hexagon on an hcp hollow-site on a slab containing 7 Cu(111) layers, was
fully relaxed until all forces on the molecule, the adatom placed on the reverse
side of the slab, and the surface layer, were smaller than 0.02 eV/{\AA}. Based
on this configuration and the molecular orientations derived from the experiment
we constructed the supercells shown in Fig.\ 3c for several different electrode
separations. These supercells were not relaxed; forces up to a few eV/{\AA} were
thus tolerated.

The electronic structure from SIESTA was used to calculate the transport
properties for the TranSIESTA\cite{Brandbyge.02.S} setup. The zero-bias conductance $G=$
G$_0T(E_F$) was derived from the transmission function $T(E)$ calculated by
Green's function techniques involving the Kohn-Sham Hamiltonian in the
scattering region and self-energies representing atomistic, semi-infinite
electrodes obtained from separate calculations for bulk Cu(111). The
transmission was sampled over a $6\times6$ $\mathbf{k}_{||}$-mesh for the 2D
Brillouin zone. To check the parameters described above, we carried out
calculations for a representative structure with increased number of $\mathbf{k}$-points
for both the electronic structure part as well as for the transmission. The
variations in the zero-bias conductance were $\leq 6$\%. The eigenchannel
visualizations, shown in Fig.\ 4a, were calculated at the Fermi energy $E_F$ for the
$\Gamma$-point according to the scheme presented in Ref.\
\onlinecite{Paulsson.07}. As seen in Fig.\ 4b the transmission functions vary
smoothly over the energy scale of the experimentally applied voltages. This justifies
the comparison with the calculated zero-bias conductances.

\textbf{Elastic response of the \buck--\buck\ junction.}.
The theoretical data in Fig.\ 3c show that the conductance of the 
\buck--\buck\ junction depends exponentially on the intermolecular separation,
approximately as $G(z)\propto e^{-2.6 \mathrm{\AA}^{-1} z}$. To prove that
the conductance does not depend significantly on the molecule--substrate distance,
we varied the electrode separation by $\pm 0.50$ {\AA} for a fixed (realistic) 
\buck--\buck\ distance of $z=9.54$ {\AA}, only to find the conductance changed
by less than 9\%. In comparison, varying the intermolecular separation
by the same amount, the conductance changes by a factor 3.7. The experimental
observation of a plateau after contact formation is therefore related to
an elastic deformation of the junction, driven by a
strong intermolecular repulsion between the two molecules.

\begin{figure}
  \includegraphics[width=0.95\linewidth]{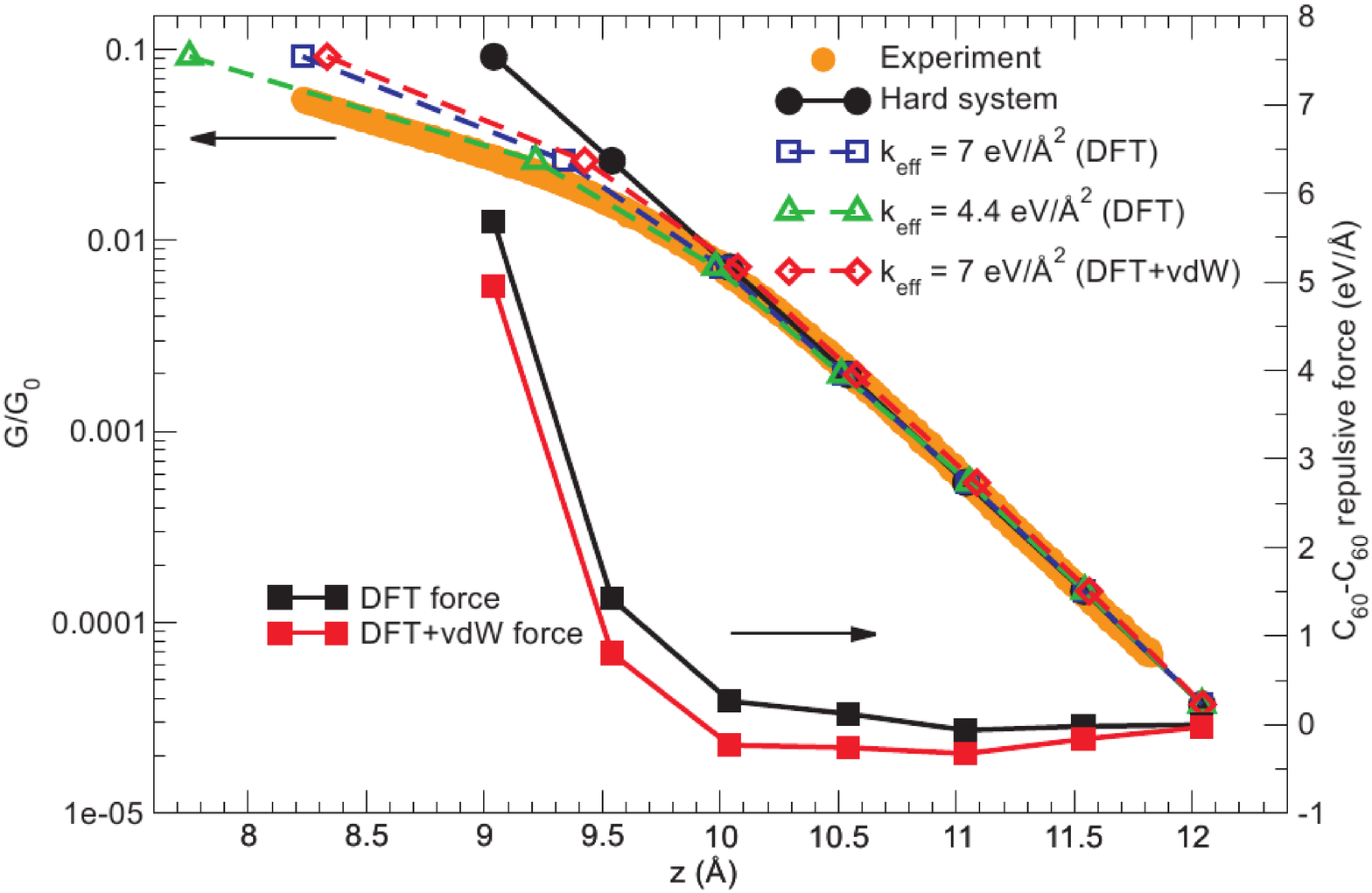}
  \caption{Conductance traces and intermolecular forces for 
the \buck--\buck\ junction with different estimates of the elasticity of the system 
and the \buck--\buck\ repulsion (with 
and without van der Waals contributions).}
  \label{figS1}
\end{figure}

To estimate the \buck--\buck\ repulsion in our structures we determined the 
average force $(F_z^{(2)}-F_z^{(1)})/2$, where $F_z^{(i)}$ is the $z$-component of 
the sum of atomic forces on molecule $i$ from our DFT calculations. Taking 
the setup with the largest intermolecular separation as a point of reference,
the evolution of intermolecular repulsion within DFT is represented in Figs.\  3c
and \ref{figS1}. For an elastic system, characterized by the effective elastic constant 
$k_\mathrm{eff}$ arising from a series of springs ($1/k_\mathrm{eff}=\sum_i1/k_i$), 
the STM piezo distances can thus be related to intermolecular distances via 
$z_\mathrm{piezo} = z_\mathrm{inter}-F/k_\mathrm{eff}$.
One example of such coordinate
renormalization is shown in Fig.\  3c where only the two soft molecule--surface segments 
were included ($k_\mathrm{eff}=k_\mathrm{m-s}/2$). One can refine this description by including 
\buck\ compressions ($k_{\mathrm{mol}}\sim 108$ eV/{\AA}$^2$ for each \buck) and 
the compliance of the metal surfaces ($k_\mathrm{surf}\sim30$ eV/{\AA}$^2$ for each 
electrode), which leads to $k_\mathrm{eff}=4.4$ eV/{\AA}$^2$ and 
an a very good agreement 
with the experimental data, cf.\ triangles in \Figref{figS1}.

While short-range interactions are generally well-described in DFT with
standard local approximations for XC, the long-range van der Waals (vdW) interactions 
are not easily incorporated in DFT because of its nonlocality. Therefore, to get an
estimate of the vdW contributions to the forces in the \buck--\buck\ contact (case 3
in Fig.\ 3), we applied an empirical correction following 
Ref.\ \onlinecite{Ortmann2006.S}. In this scheme the interaction energy between each pair of
atoms ($i$, $j$), separated by the distance $R$, is corrected by an additional attractive
energy the form $U_{ij}^\mathrm{vdW}=-f_D^{ij}(R)C_6^{ij}/R^6$, where
$f_D^{ij}(R)=1-\exp[-0.00075 R^8/(r_i+r_j)^8]$ is a damping function that removes
contributions for small separations. The parameter $r_i$ is a characteristic covalent radius of
atom $i$. We used $r_\mathrm{C}=0.7245$ {\AA} (corresponding to the average bond length in our
description of \buck) and $r_\mathrm{Cu}=1.308$ {\AA} (derived from the Cu lattice
constant). For the $C_6$ coefficient we used 
the London formula $C_6^{ij}=\frac 32 \alpha_i\alpha_j I_iI_j/(I_i+I_j)$ with 
experimental values \cite{CRChandbook} for the polarizability $\alpha_i$ and the 
ionization potential $I_i$ of atom $i$. The strategy outlined above enabled us to reproduce 
the 3D fcc crystal of \buck\ with a lattice constant of 10.02 {\AA} and a bulk modulus only
about 12\% lower than the experimental value. Applying the vdW correction to the \buck--\buck\ contact
the forces shown in \Figref{figS1} are obtained. Since the intermolecular repulsion
is only slightly reduced by the attractive vdW forces, the coordinate renormalization
of the conductance trace is mostly sensitive to the elastic constant $k_\mathrm{eff}$ and not to the force estimate.
Furthermore, since it is generally argued that the London form shows a tendency to overestimate
the vdW interaction [see Ref.~\onlinecite{Ortmann2006} and references herein], we conclude that
vdW contributions play a minor role for understanding the experimental conductance trace.

\end{document}